\documentclass[12pt]{article}
\usepackage{graphicx}
\usepackage{epstopdf, epsfig}
\pdfoutput=1

\usepackage[a4paper, margin=2.5cm]{geometry}
\usepackage{amsmath} % For equation numbering
\usepackage{multirow} % To combine multiple rows
\usepackage{amssymb}
\usepackage{natbib}
\usepackage{caption}
\usepackage{bm}
\usepackage{tabularx}
\usepackage{mathtools}

\usepackage{placeins} % For float barriers
\usepackage{hyperref}
\usepackage[mathscr]{euscript}
 \let\mathscr\relax
\usepackage[scr]{rsfso}

\usepackage{enumitem}

\usepackage{cleveref}
\crefformat{section}{\S#2#1#3}

\title{Finding a reconnection in a vortex-haystack}

\author{\small Abhishek Harikrishnan$^1$, Benjamin Musci$^1$, Adrien Lopez$^1$, Bartosz Protas$^2$, Adam Cheminet$^1$, \\ \small  Jean Le Bris$^1$, Jean-Marc Foucaut$^3$, Jean-Philippe Laval$^3$,  Christophe Cuvier$^3$, \\ \small Pierre Bragança$^3$,  B{\'e}reng{\`e}re Dubrulle$^1$}

\date{\small 
$^1$CEA, CNRS, SPEC, Université Paris-Saclay, Gif sur Yvette, France, \\ $^2$Department of Mathematics and Statistics, McMaster University, Hamilton, Canada, \\ $^3$Univ. Lille, CNRS, ONERA, Arts et Metiers ParisTech, Centrale Lille, France}

\begin{document}

% ------------------- TITLE ------------------- %

\maketitle

% ------------------- ABSTRACT ------------------- %

\begin{abstract}

\noindent Counter-rotating vortex reconnection is a topology-transforming event that leads to strong vorticity amplification and is often associated with the most irregular regions of the velocity field. Detecting such events directly in turbulent flows is a major challenge due to the ambiguity inherent in the definition of a vortex and the rapid proliferation of intense small scale vortices. We develop a novel methodology by combining two tools. The first is a local in scale, space and time regularity indicator that isolates spatio-temporal regions of interest. The second is a reconnection indicator that exploits a well-known property of vortex reconnection namely, that the enstrophy (volume integral of the square of vorticity) flows from the symmetry to the collision plane. To examine the flow of enstrophy in the correct reference frame, a new method inspired by orientational order parameters in liquid crystals is developed. We then show a proof of concept for our reconnection indicator by comparing it with diagnostic methods available in the literature. The method is validated on experimental datasets from a turbulent von K{\'a}rm{\'a}n flow at two different resolutions, one of which allows us to probe the flow below the Kolmogorov scale, $\eta$. Our findings suggest that counter-rotating vortex reconnection could be a potential mechanism for the creation of motions at scales below $\eta$.

\end{abstract}

% ------------------- KEYWORDS ------------------- %
% This should be removed at the end of writing the paper.
% Author will add this while submission
% They will be included in the typesetting process.

%\begin{keywords}
%	Atmospheric Boundary Layer, Stratification, Coherent Structures, Global intermittency
%\end{keywords}

% -------------------1. INTRODUCTION ------------------- %

\newpage

\section{Introduction}
\label{sec: Introduction}

\begin{figure*}[t!]
\centering
\includegraphics[width=17.8cm]{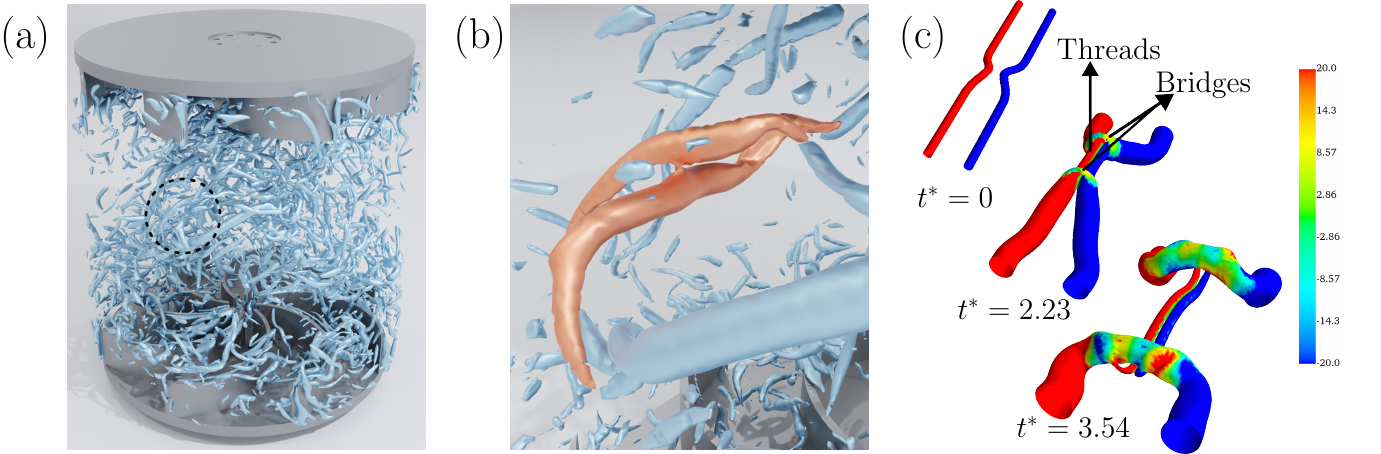}
\caption{(a) Q-criterion isosurfaces computed from the DNS of \citet{faller2021nature}. (b) A zoom of the the circled region in (a). The red structures are undergoing reconnection as identified by \citet{gfm2022}. (c) DNS of vortex reconnection showing the formation of bridges and threads. Q-criterion isosurfaces, colored with axial vorticity, are thresholded at $>0.01$Q$_{\max}$. The non-dimensional time $t^* = t\Gamma/2\pi(2s_0)^2$ where $\Gamma$ is the initial circulation strength and $2s_0$ is the initial separation distance of the vortex tubes. Further details of this DNS simulation are provided in Section 1 of the supplementary.}\label{figure: reconnection}
\end{figure*}

Turbulent flows are characterized by the presence of swirling motions, extending towards the smallest scales as intense vortex tubes with a typical width of a few Kolmogorov scales ($\eta$) and typical velocity gradients that are a few orders of magnitude larger than the mean \citep{yeung2015extreme, buaria2019extreme, buaria2026turbulence}; see also Fig.~\ref{figure: reconnection}(a).
These structures are often considered the building blocks of turbulence, or as stated by \citet{kuchemann1965report}  --- the ``sinews and muscles of fluid motions".
Empirical observations indicate that as the viscosity decreases, these initially circular vortices become flatter upon approaching one another and eventually generate intense vorticity at the point of contact \citep{yao2022vortex, harikrishnan2025viscous}. 
This triggers questions about their fate in the asymptotic limit when viscosity approaches zero, or equivalently, as the Reynolds number $Re\to \infty$. 
This question has important mathematical implications, as it is intimately linked with the issue of well-posedness and blow-up of the fluid equations. 
In particular, theoretical models of inviscid interaction of these vortices based on the Biot-Savart approximation indicate that they may result in a `physical' singularity with diverging vorticity \citep{moffatt2019towards1, moffatt2019towards2}. 
However, direct numerical simulations (DNS) of vortex tube interaction in the presence of viscosity point to an alternate scenario where the vortex tubes undergo a reconnection process. Here, the outermost vortex lines of each of the approaching vortex tubes are annihilated by viscous cross-diffusion and bridge structures are formed in a direction orthogonal to their initial orientation --- which curbs the growth of vorticity \citep{yao2020physical}; see also Fig.~\ref{figure: reconnection}(c). 
However, the peak of vorticity and enstrophy (volume integral of the square of vorticity) continues to increase with increasing $Re$ and whether this leads to an actual blow-up in the inviscid limit is still an open question. 

Even with recent computational advances, it is difficult to answer this question with DNS as the computational burden increases $\propto$ $Re^{3}$.
Laboratory experiments, on the other hand, can be used to achieve relatively large $Re$.
While vortex reconnection has been experimentally observed in viscous fluids \citep{mckeown2018cascade, lim1992instability}, superfluids \citep{fonda2019reconnection, stasiak2025experimental} and optical waves \citep{adams2024spatiotemporal}, the primary experimental issue is the difficulty in measuring the required fluid quantities with sufficient spatio-temporal resolution to capture potential singularity mechanisms. 
For the swirling von K{\'a}rm{\'a}n flow studied here, the Kolmogorov length scale is related to the dimensionless energy dissipation rate ($\epsilon^*$) by $\eta = R(Re^3 \epsilon^*)^{-1/4}$ and $R$ is the tank radius \citep{debue2019experimental}. 
This linear relationship between $R$ and $\eta$ allows us to use a Giant von K{\'a}rm{\'a}n (GvK) facility with large $R = 0.5$m to simultaneously achieve high $Re$ and a uniquely high experimental resolution.
We analyze GvK datasets that feature two different $Re$ and resolved scales, as shown in Table \ref{table: GvK parameters}. 
To the authors' knowledge, the case C1 ($Re = 6\,269$), allows for the first time, a time-resolved 3D experimental exploration of scales smaller than $\eta$ at $l = 0.76 \eta$.
Further details on the experiment can be found in Appendix  \hyperref[subsection: GvK]{A}.

While the reconnection mechanism and its intricate details are generally studied by creating vortex filaments \citep{siggia1985collapse} or vortex tubes with gaussian or similar velocity profiles \citep{yao2020physical}, detecting such events directly in a turbulent flow is akin to finding a needle in a haystack (cf. Fig.~\ref{figure: reconnection}(a, b)). The goal of this paper is to show that it is possible to systematically identify reconnection events in a fully turbulent flow with the following two-step procedure. 

First, spatio-temporal regions of interest are isolated by identifying extreme events using a scale-dependent regularity indicator, $D_l^I(\bm{x})$, which is inspired by mathematical work on energy conservation in singular solutions of the Navier-Stokes equations; see Section \ref{subsec: DR_indicator}. Second, within the time-window of the identified extreme event, an enstrophy-based reconnection criterion is applied \citep{kang2020maximum}. 
This criterion, described in Section \ref{section: reconnection_criterion}, exploits a well-known property of reconnection whereby enstrophy flows from the symmetry to the collision plane (see Fig.~\ref{figure: n_tensor_tests} for a description of these planes). This allows one to identify a reconnection event when the ordering of the component-wise enstrophies corresponding to three  orthogonal coordinate directions (e.g., the Cartesian directions $x$, $y$, $z$) changes. 
While this method works when the vortex tubes are aligned with known coordinate directions, it does not translate well to turbulent flow fields where the appropriate reference frame to assess the flow of enstrophy is a priori unknown and continually changing. 
Thus, we introduce a novel methodology inspired by orientational order parameters for phase transitions in liquid crystals to find an optimal reference frame where vorticity shows the strongest alignment with the new coordinate directions. 
These tools are then applied to the experimental datasets (Section \ref{sec: results}), providing a proof-of-concept for our methodology. 

\begin{table}[t!]
\centering
\caption{Experimental parameters for the two GvK datasets. The Reynolds number $Re = 2\pi R^2 F/\nu$, where $R$ is the tank radius, $F$ is the frequency of rotation of the impellers and $\nu$ is the kinematic viscosity. $\langle \epsilon\rangle$ is the average dissipation rate and $\eta$, $\tau_{\eta}$ are the Kolmogorov length and time scales respectively.}
\begin{tabular}{ccccccc}
\hline
Case & Re & Runs & F [Hz] & $\langle \epsilon \rangle$ [m$^2$s$^{-3}$] & $\eta$ [mm] & $\tau_{\eta}$ [ms] \\
\hline
C1 & $6\,269$ & $40$ & $0.004$ & $6.61\times10^{-8}$ & $1.98$ & $3893$\\
C2 & $156\,719$ & $36$ & $0.1$ & $1.03\times10^{-3}$ & $0.18$ & $31$\\
%\bottomrule
\hline
\end{tabular}
\label{table: GvK parameters}
\end{table}

\section{Description of the tools}
\label{sec: description}
\subsection{Regularity indicator}
\label{subsec: DR_indicator}
One must first isolate regions of interest by identifying extreme events of `inertial' energy dissipation. 
This quantity stems from the seminal paper of \citet{duchon2000inertial}, who derived an energy balance for weak solutions of the incompressible Navier-Stokes equations (INSE),
\begin{equation}
    \partial_t \frac{u_i^2}{2} + \partial_j\bigg[u_j\bigg(\frac{u_i^2}{2} + p\bigg)\bigg] = \nu\Delta\frac{u_i^2}{2} - \nu\partial_j u_i \partial_j u_i + D(\bm{u}).
\end{equation}
Here, $\bm{u}$ is a weak solution of INSE and its derivatives are understood in the sense of distributions, whereas
$D (\bm{u})$ is an additional term to the regular energy balance of 3D INSE.
It quantifies the energy being dissipated by non-viscous means which arises due to lack of smoothness of the velocity field. 
The goal of \citet{duchon2000inertial} with this energy balance was to illustrate Onsager's conjecture \citep{onsager1949statistical} who first proposed such a dissipation mechanism. To make it usable for experimental or numerical datasets, a coarse-grained version of this quantity defined as
\begin{equation}
    D_l^I[\bm{u}](t,\bm{x}) = \frac{1}{4} \int \nabla \phi^l(\bm{\xi})\bm{.}\delta{\bm{u}}(\delta\bm{u})^2 d^3\bm{\xi}
\end{equation}
is used, where $l$ is the probed scale \citep{dubrulle2019beyond}. 
At $l = \eta$, one expects $D_l^I \rightarrow 0$ as all injected energy should be dissipated by viscosity. 
It should also be noted that $\lim_{l\to0}D_l^I(\bm{u}) = D(\bm{u})$.
A Gaussian filter $\phi^l (\bm{x}) = \exp(-30\bm{x}^2/2l^2) / (2\pi l^2/30)^{3/2}$ is used to smooth the velocity field and $\delta\bm{u} (\bm{x, \xi}) = \bm{u}(\bm{x} + \bm{\xi}) - \bm{u}(\bm{x})$ is the velocity increment that depends on an increment distance $\bm{\xi}$. 
%
% \begin{equation}
%     \lim_{l\to0}D_l^I(\bm{u}) = D(\bm{u})
% \end{equation}
%
Thus, $D_l^I$ can be interpreted as a local (in space, scale and time), indicator of energy transfer with positive and negative values indicating downscale and upscale energy transfers, respectively. 
At $l \leq \eta$, if $D_l^I (\bm{u}) = 0$, the velocity field is sufficiently regular.  
Conversely, when $D_l^I (\bm{u}) \gg 0$ it corresponds to spatio-temporal locations that are irregular at scale $l$, as noted by \citep{dubrulle2019beyond}.
Therefore, by simultaneously decreasing $l$ as much as possible while considering the extrema of $D_l^I$, one identifies the flow field locations with lowest regularity.

Previous studies have shown that extreme $D_l^I$ events are associated with predominantly vortex-like structures \citep{saw2016experimental, debue2021three, musci2025experimental}. 
Furthermore, \citet{cheminet2022eulerian} showed with a Lagrangian irreversibility indicator (which can be thought of as a combination of $D_l^I$ and a viscous dissipation term) that the most irreversible trajectories tend to be at the intersection of two vortices, suggesting a link between extreme events of $D_l^I$ and \textit{vortex reconnection} which is leveraged in this work. 

\begin{figure}[t!]
\centering
\includegraphics[width=0.7\linewidth]{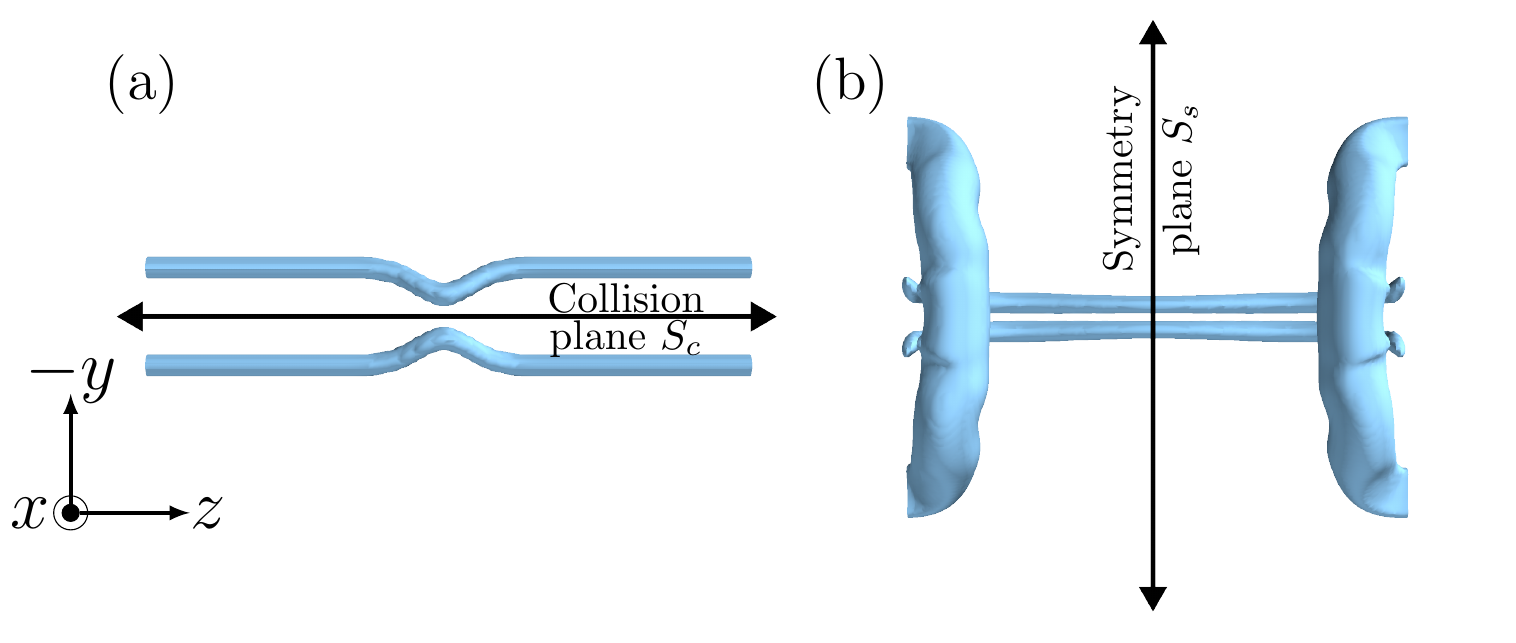}
\caption{Isosurfaces of Q-criterion structures thresholded at Q$>0.01$Q$_{\max}$ are shown here (a) before ($t^* = 0$) and (e) after reconnection ($t^* = 4.48$). Also shown are the symmetry (x-y) and collision (x-z) planes. Black arrow indicates the direction of the eigenvector of N-tensor corresponding to the largest eigenvalue.}\label{figure: n_tensor_tests}
\end{figure}

 \subsection{Reconnection indicator}
 \label{section: reconnection_criterion}
 
Apart from identifying regions of interest, a primary challenge regarding the detection of vortex reconnection directly in turbulent flows is that there is no generally accepted mathematical definition of a vortex \citep{haller2005objective}. 
\citet{gunther2018state} provides a wide range of criteria that are currently being employed in the literature. 
The diagnostic used here for the visualization of vortex-like features is the popular Q-criterion \citep{hunt1988eddies} which defines vortices as regions where rotation dominates over the strain, i.e.,
\begin{equation}
    \text{Q} = \frac{1}{2} (\| \Omega \|^2 - \| S \|^2) > \text{Q}_{\text{th}}.
    \label{eq:Q}
\end{equation}
Here, $\Omega = \frac{1}{2} [\nabla\bm{u} - (\nabla\bm{u})^T]$ is the vorticity or spin tensor, $S = \frac{1}{2} [\nabla\bm{u} + (\nabla\bm{u})^T]$ is the strain-rate tensor, $\nabla\bm{u}$ is the velocity gradient tensor, $\| \cdot \|$ is the Frobenius norm and $\text{Q}_{\text{th}}$ is some arbitrary threshold. 
If the Q-criterion, or any of its related methods (see Section 5.1 of \citet{gunther2018state}), are used to detect vortex reconnection, then the user's choice of the threshold dictates if/when a reconnection takes place. 
While Lagrangian techniques overcome the threshold barrier, the integration time to advect tracers and methodology to extract features are often unclear.
Techniques such as the Lagrangian Averaged Vorticity Deviation (LAVD) and related, Instantaneous Vorticity Deviation (IVD) overcome these issues but as pointed out by \citet{neamtu2025unsupervised} several parameters tend to be user-defined, need tuning, and may require apriori knowledge of the structures' location and orientation. 
Therefore, a technique that does not rely on the definition of a vortex is needed. 

As described by \citet{melander1988cut}, during a counter-rotating vortex reconnection, the vortex tubes are pushed close to each other due to curvature induced self-induction and mutual induction (when they are sufficiently close). 
Since the tubes have opposite circulation, it is annhilated at the point of contact due to viscous cross-diffusion and the circulation is transferred continuously from the symmetry plane $S_s$ to the collision plane $S_c$ resulting in the formation transverse vortex structures called bridges.
Therefore, the transfer of circulation offers a key evidence of the reconnection process that can be leveraged. 
However, a primary issue with using circulation is that it is calculated only at a single plane along the vortex tube which cannot be chosen unambiguously in a fully turbulent flow.
\citet{kang2020maximum} noted that there is a transfer of enstrophy among its different components during reconnection. Note that component-wise enstrophy is defined as,
\begin{equation}
    \mathcal{E}_i (\bm{u}(t)) \coloneqq \int_{\Omega} |(\bm{\nabla} \times \bm{u}(t)) \cdot \bm{e}_i |^2 d\bm{x}, \quad \bm{i} = 1, 2, 3,
\end{equation}
where $\Omega$ is the integration domain and $\bm{e}_i$ are unit basis vectors of an orthogonal coordinate system. This quantity is preferred over circulation since it is calculated over the entire domain representing the complete vortex tubes and is a more forgiving metric independent of the sign. 

In all earlier studies the authors took $\bm{e}_i$ to be the unit basis vectors of the global Cartesian coordinate system. 
This is sufficient for vortex tubes that are already aligned along one of the coordinate axis as shown in Fig. \ref{figure: n_tensor_tests}.
However, in a fully turbulent flow, the vortex tubes can be aligned along any direction and one must find a more suitable reference frame to detect the transfer of enstrophy.

To do this, we draw inspiration from the liquid crystal (LC) theory. Since the phase of LC depends on the orientation of the molecules, any theoretical treatment of LC involves an analysis of the orientational order. 
This is generally  accomplished with a second-rank, traceless, symmetric tensor called the `Q-tensor.' 
To avoid confusion with the Q-criterion used for vortex detection, cf.~\eqref{eq:Q}, we will refer to the Q-tensor as the Nematic, or N-tensor ($N_{ij}$). 
For uniaxial LC, 
\begin{equation} \label{eq:Q_tensor}
N_{ij} = \left( n_i n_j - \frac{1}{d} \delta_{ij} \right)
\end{equation}
where $\bm{n}$ is the director which is a dimensionless unit vector that represents the preferred orientation of molecules in the neighborhood of any point, $\delta_{ij}$ is the Kronecker delta and $d$, the number of dimensions. 
Similar to the molecules of LC, we can consider that a vortex tube is composed of a collection of vorticity vectors oriented along some direction. 
If the director $\bm{n}$ is replaced by vorticity $\boldsymbol{\omega}$, 
\begin{equation} \label{eq:Q_tensor_vorticity}
N_{ij} = \left( \omega_i \omega_j - \frac{1}{D} |\boldsymbol{\omega}|^2 \delta_{ij} \right)
\end{equation}
and $|\cdot|$ is the Euclidean norm, this modified N-tensor can then describe the local alignment of a vortex tube with the coordinate directions (up to a sign). 
The behavior of the N-tensor can be illustrated with a 2D analysis involving a single vorticity vector. 
The N-tensor in the matrix form is then
\begin{equation} \label{eq:Q_tensor_2D_oneVec}
N_{ij} = 
\begin{bmatrix}
\omega_x^2 - \frac{1}{2} |\boldsymbol{\omega}|^2 & \omega_x \omega_y\\
\omega_y\omega_x & \omega_y^2 - \frac{1}{2} |\boldsymbol{\omega}|^2
\end{bmatrix}
\end{equation}
where $\boldsymbol{\omega} = [ \omega_x, \omega_y]^T$. Thus, the eigenvalues and their corresponding eigenvectors are
\begin{equation*}
    \lambda_{1, 2} = \pm \frac{1}{2} |\boldsymbol{\omega}|^2, \quad v_1 = \begin{bmatrix} \omega_x \\ \omega_y \end{bmatrix}, \quad v_2 = \begin{bmatrix} -\omega_y \\ \omega_x \end{bmatrix} \, .
\end{equation*}
The key observation is that the eigendirection defined by the eigenvector $\bm{v}_{\lambda_{\max}} = \bm{v}_1$ associated with the largest eigenvalue $\lambda_{\max} = \lambda_1$ is aligned with the vorticity vector $\boldsymbol{\omega}$ and this remains the case when the vorticity vector is rotated by some arbitrary angle $\theta$, i.e., 
\begin{equation*}
    \omega = \begin{bmatrix} \cos \theta \\ \sin \theta \end{bmatrix},\quad \lambda_{1, 2} = \pm \frac{1}{2}, \quad v_1 = \begin{bmatrix} \cos \theta \\ \sin \theta \end{bmatrix}, \quad v_2 = \begin{bmatrix} -\sin \theta \\ \cos \theta \end{bmatrix}.
\end{equation*}
For a large number of vorticity vectors within a volume $\Omega$, one can simply average the N-tensor over space,
\begin{equation} \label{eq:Q_tensor_vorticity_averaged}
\langle N_{ij} \rangle = \frac{1}{\Omega} \int_{\Omega} N_{ij} (\bm{x}) \; d\bm{x}
\end{equation}
% and $S$ is the scalar order parameter. Increasing $S$ from $0 \to 1$ changes the molecular alignment with the director from being completely random to perfectly aligned. Here, $S$ is set to $1$ to ensure perfect alignment of vorticity vectors with the vortex tube.
and then the eigenvector $v_{\lambda_{\max}}$ represents the prevalent directions of the vorticity vectors in the region $\Omega$.  The alignment is also correctly determined for antiparallel vortex tubes where the vorticity vectors in each tube are opposite to each other. 
Repeating the 2D analysis with two such antiparallel vorticity vectors, the eigenvalues and eigenvectors of the averaged N-tensor are
\begin{equation*}
    \omega_1 = \begin{bmatrix} 1 \\ 0 \end{bmatrix}, \omega_2 = \begin{bmatrix} -1 \\ 0 \end{bmatrix}, \quad \lambda_{1, 2} = \pm \frac{1}{2}, \quad v_1 = \begin{bmatrix} 1 \\ 0 \end{bmatrix}, \quad v_2 = \begin{bmatrix} 0 \\ 1 \end{bmatrix}
\end{equation*}
where $\bm{v}_{\lambda = \lambda_{\max}}$ is again aligned with the direction of the vorticity vectors.
In an actual reconnection scenario, $\bm{v}_{\lambda_{\max}}$ lies in the symmetry and collision plane directions before and after the reconnection, respectively, as shown in Fig. \ref{figure: n_tensor_tests}. 

\begin{figure}[t!]
\centering
\includegraphics[width=0.7\linewidth]{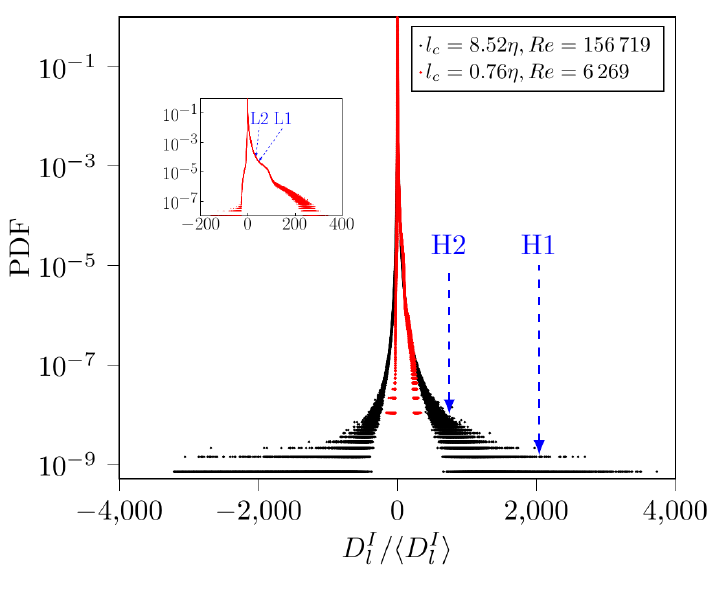}
\caption{Probability density function (PDF) of mean-normalized inertial energy dissipation for the two datasets described in table \ref{table: GvK parameters}. Here, $\langle \cdot \rangle$ denotes the space-time average. The inset highlights the smallest probed scale, case C1, $l = 0.76\eta$ at $Re = 6\,269$. Two extreme events each from case C1 denoted L1, L2 and from case C2, denoted H1 and H2 are examined.}\label{figure: pdf DR}
\end{figure}

Prior to reconnection, if the eigenvalues are ordered as $\lambda_1 \geq \lambda_2 \geq \lambda_3$, their corresponding eigenvectors will define a new reference frame represented by the matrix $\mathcal{V} = \left[ \bm{v}_1, \bm{v}_2, \bm{v}_3 \right]$ where $\bm{v}_1$ points in the symmetry plane. 
Taking the dot product with the vorticity vector $\bm{\omega}_{\mathcal{V}} = \bm{\omega}^T\cdot\mathcal{V}$ projects the vorticity onto the new reference frame. 
Therefore, at initial time, $\mathcal{E}_1 \geq \mathcal{E}_2 \geq \mathcal{E}_3$. 
Then it suffices to check if $\mathcal{E}_2 > \mathcal{E}_1$ or $\mathcal{E}_3 > \mathcal{E}_1$. 
In other words, a reconnection event is detected when there is a reordering of the component-wise enstrophy due to the flow of enstrophy from the symmetry to the collision plane, which can be associated with either $\mathcal{E}_2$ or $\mathcal{E}_3$.

\section{Application to experimental data}
\label{sec: results}

\subsection{Identifying extreme events of inertial energy dissipation} 
\label{subsec: extreme events of DR}
Probability density functions (PDFs) of the mean-normalized $D_l^I$ are shown in Fig.~\ref{figure: pdf DR} for the two Reynolds numbers and their corresponding probed scales. Similar to the observations of \citet{musci2025experimental}, the PDFs are non-Gaussian with right-skewed tails, indicating strong and predominantly downscale energy transfer events. 
The tails for larger Reynolds numbers, and correspondingly larger probed scales, reveal events thousands of times the mean value in strength, consistent with the descriptions of ``extreme events'' in turbulence literature \citep{schumacher2010extreme, yeung2015extreme}. 
The inset shows the case with the smallest probed scale below $\eta$. 

\begin{figure}[t!]
\centering
\includegraphics[width=\linewidth]{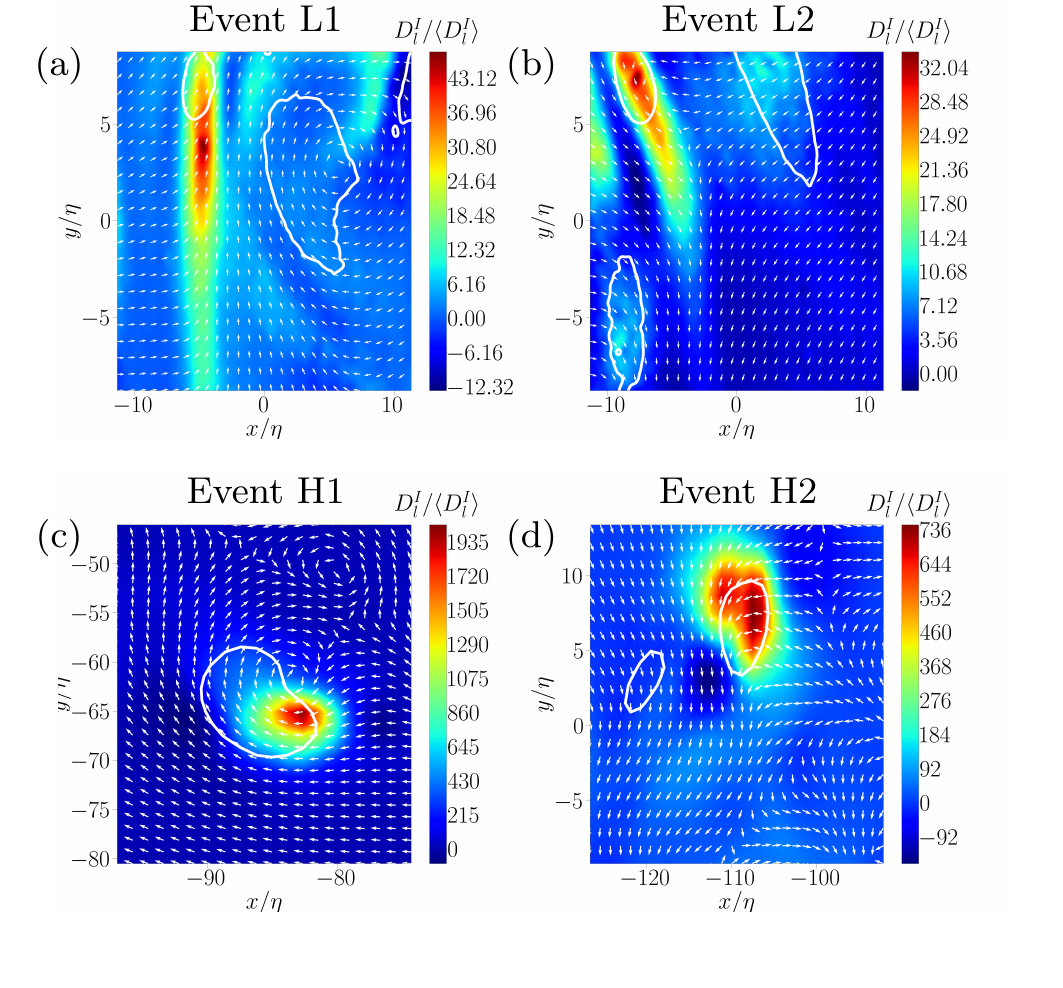}
\caption{Spatial distributions of $D_l^I / \langle D_l^I \rangle$ in (a, b) case C1 and (c, d) case C2 on a subset of the domain. Thick white lines are contours of Q-criterion thresholded at $0.4, 0.39, 0.16$ and $0.34$ Q$_{\text{max}}$ respectively. White arrows indicate the velocity field around the events. }\label{figure: DR plane}
\end{figure}

To detect the extreme events of $D_l^I$, the spatial maxima of this quantity are calculated at each timestep for all experimental runs.
The strongest event of every run is then identified and sorted. 
A typical time series for a full run is shown in Fig.~\ref{figure: combined vortex plot}(d). 
The portion highlighted in red is an extreme event that lasts roughly $3.5$ Kolmogorov time scales ($\tau_\eta$) showing periods of steep growth and decay. 
Examining $10$ strongest extreme events at $l = 0.76\eta$ ($Re = 6\, 269$, case C1) suggests that they last about $3.2\tau_\eta$ on average, with $1.2\tau_{\eta}$ and $5.2\tau_{\eta}$ being the lifetime of the shortest and the longest events, respectively. 
Repeating this at $l = 8.5\eta$ ($Re = 156\, 719$, case C2), yields a similar mean of $2.5\tau_{\eta}$, with $0.5\tau_{\eta}$ and $6.5\tau_{\eta}$ being the lifetime of the shortest and longest events, respectively. 
These results are in line with \citet{yeung2015extreme} who reported extreme events of enstrophy and dissipation in a DNS of isotropic turbulence at $Re_{\lambda} \approx 1300 $ that lasted about $3-5\tau_{\eta}$.

Once the extreme events are detected temporally, their spatial characteristics are examined. Two extreme events, one from each of the cases C1 and C2 (cf. Fig. \ref{figure: pdf DR}), are visualized in Fig.~\ref{figure: DR plane}. 
The spatial distribution of $D_l^I / \langle D_l^I \rangle$ is dominated by positive values, indicating downscale energy transfer even below the Kolmogorov scale for events L1 and L2. The thick white lines are contours of Q-criterion that demarcate vortex-like regions around the event. Similar to the observations of \citet{dubrulle2019beyond}, it can be seen that the maxima of $D_l^I / \langle D_l^I \rangle$ are located near or within a Q-criterion structure. 

\subsection{Detecting vortex reconnection in turbulent flow data}
\label{subsec: detecting vortex reconnection}

To examine if a reconnection has taken place, we first resort to diagnostic methods available in the vortex-reconnection literature. 
Through several of their works, \citet{yao2020physical, yao2022vortex} and references therein, have distilled telltale signs of antiparallel (or counter-rotating) vortex reconnection. 
These include: 
A) a sharp growth and subsequent decay of the total enstrophy $\mathcal{E}$ and of the supremum norm of the vorticity $\| \boldsymbol{\omega} \|_{\infty}$,  
B) a strong growth and subsequent decay of the net absolute helicity when the vortex tubes are weakly polarized, i.e., there is a weak axial flow in both vortex tubes; here, the net absolute helicity is defined as $\mathcal{H}_a = \int_{\Omega} |h| d\Omega$ where $h = \bm{u} \cdot \bm{\omega}$, 
C) the separation distance between the two vortices during approach that follows the scaling $s_0 \sim (t_0 - t)^{0.5}$ ($t_0$ is the time at which the vortices reconnect) %for thin, viscous vortex cores and larger exponents for thicker ones 
and finally, 
D) null vorticity at the point of contact. 
Other visual flow diagnostics include flattening of the vortex cores during approach and formation of bridge and thread structures. 

These diagnostics are applied to all four events shown in Fig.~\ref{figure: DR plane}. 
For brevity, only the event L2 is discussed in the running text of the paper and the others can be found in the supplementary material. 
Fig.~\ref{figure: combined vortex plot}(d) shows the full time series for this run and suggests that this event lasts approximately $3.5 \tau_{\eta}$ time units. The event time window highlighted red is used for all subsequent analysis. 

Examining the visualization at $t = 0.5\tau_{\eta}$ (cf. Fig. \ref{figure: combined vortex plot}(a1)), one can see that two large Q-criterion vortex structures exist. 
For the chosen threshold (Q $> 0.06$Q$_{\text{max}}$) the structure shaded blue is of size $11.9\eta \, \times \, 12.6\eta \, \times \, 3.2\eta$ whereas the red one is $5.5\eta \, \times \, 6.5\eta \, \times \, 3.2\eta$. 
Although only a very short extent is available in the $z$-direction, a rather well-known feature of reconnection is that, regardless of their initial orientation, the structures tend to become antiparallel at the point of contact, and therefore, should not hinder with the detection \citep{siggia1985incipient, yao2022vortex}.
By coloring these isosurfaces with axial vorticity (in this case, $\omega_z$ as the vortices are roughly oriented in the z-direction), we can infer that these vortices are counter-rotating. 
The first phase of reconnection, i.e., inviscid advection is evident, as the vortices approach each other over time. 
However, due to the limited extent in the z-direction, it is difficult to verify other visual diagnostics such as flattening of the vortex cores or the formation of bridge and thread structures.
Even though a third structure is visible on the right side of the domain in Figs. \ref{figure: combined vortex plot}(a2-a3), the spatial distribution of mean-normalized $D_l^I$ (Figs. \ref{figure: combined vortex plot}(b2-b3)) confirms that the maxima is associated with the reconnection event and is not influenced by the structure that is being swept into the domain. 
The distribution of vorticity magnitude $|\boldsymbol{\omega}|$ in Figs. \ref{figure: combined vortex plot}(c2-c3) suggest that vorticity becomes small in-between the two vortices which is characteristic of a reconnection event. 

A small box of size $13.2\eta \times 19.9\eta \times 3.2\eta$ (colored red in Fig.~\ref{figure: combined vortex plot}(a2)) is chosen to exclude the influence of the large vortex seen on the right side of the domain. 
Within this box and the event time window, a steady rise and decay of both the total enstrophy and the maximum norm of vorticity can be seen in Fig.~\ref{figure: combined vortex plot}(e, f). 
The mean axial velocity ($u_z$) at the initial time of the event time window within both vortex cores has the same sign suggesting that the axial flow is in the same direction. 
This means that the helicity is preserved longer as the reconnecting vortex lines are twisted in the same direction \citep{mcgavin2019reconnection}. Therefore, one should observe a strong growth of the net absolute helicity. As seen from Fig.~\ref{figure: combined vortex plot}(g), the net absolute helicity normalized by the initial value shows a sharp rise and followed by a decrease consistent with a weakly polarized case discussed in \citet{mcgavin2019reconnection, yao2021polarized}. 
Next, the vortices are isolated and tracked in time with a simple volume overlap technique until they merge. 
The separation distance between the two vortices is found to scale as $s_0 \sim (t_0  - t)^{0.88}$ (see Fig.~\ref{figure: combined vortex plot}(h)). 
This is to be expected as the local assumption required for the $1/2$ scaling is maintained only for thin vortex cores \citep{yao2022vortex}.

While the above diagnostics indicate a possible reconnection, as stated previously the key evidence is provided by the transfer of component-wise enstrophy from the symmetry $S_s$ to collision $S_c$ planes. 
As the vortices are roughly oriented in the $z$-direction, one should expect a flow of enstrophy from the component $\mathcal{E}_3$ to either $\mathcal{E}_2$ or $\mathcal{E}_1$. 
Fig.~\ref{figure: combined vortex plot}(i) shows that there is indeed a flow of enstrophy from $\mathcal{E}_3$ to $\mathcal{E}_2$ suggesting a reconnection at $t = 1.71\tau_{\eta}$.
We emphasize that a detection without using the correct reference frame is only possible for this particular event because the vortices are roughly oriented in the $z$-direction. 
It can be seen that for the event L1 (see Section 2 in the supplementary), this detection fails without the correct reference frame even though the vortices are more or less oriented in the the $z$-direction. 

\begin{figure*}[t!]
\centering
\includegraphics[width=\textwidth]{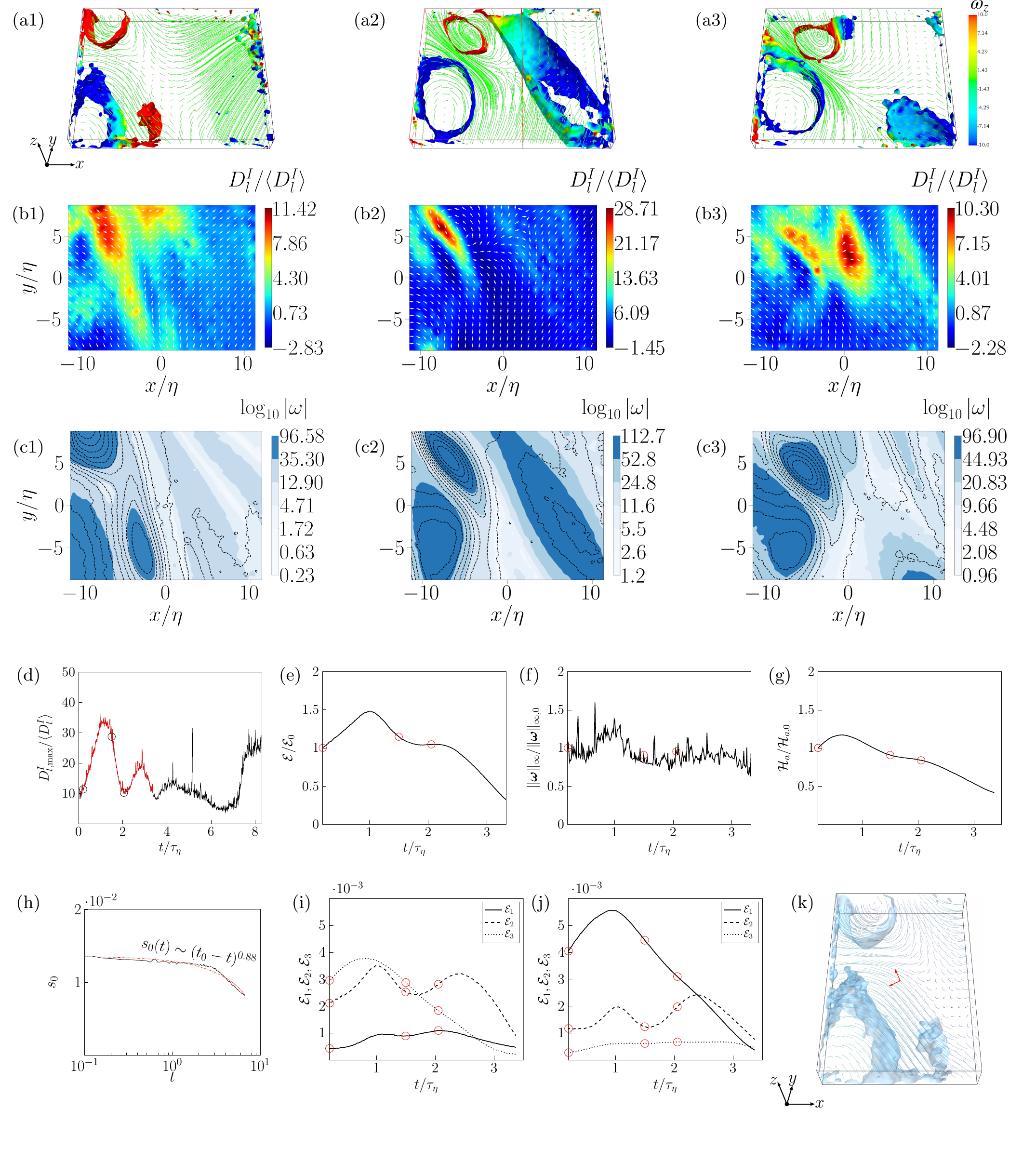}
\caption{(a1-a3) Isosurfaces of the Q-criterion thresholded at Q $> 0.06$ Q$_{\max}$ and shaded with axial vorticity ($\omega_z$). Velocity streamlines at mid-z height are shown. Horizontal slices of (b1-b3) mean-normalized $D_l^I$ and (c1-c3) vorticity magnitude are plotted at slices where the mean-normalized $D_l^I$ reaches an instantaneous maximum. All the above are plotted at times $t /\tau_{\eta} = 0.2, 1.5, 2.05$. These times are indicated in (e-g, i-j) with unfilled red markers. (d) Full time series of mean-normalized $D_l^I$ for the run. Temporal evolution of (e) enstrophy, (f) maximum norm of vorticity and (g) net absolute helicity, all of which are normalized by their initial value. Behavior of (h) separation distance, component-wise enstrophy computed (i) without and (j) with the new reference frame obtained from the N-tensor. This reference frame is indicated by red arrows in (k) and is oriented such that the eigenvector $v_{\lambda_{\max}}$ is normal to the plane of the paper.}\label{figure: combined vortex plot}
\end{figure*}

The above analysis is now repeated by calculating component-wise enstrophy in the reference frame dictated by the N-tensor (see Fig.~\ref{figure: combined vortex plot}(k)). 
Note that the N-tensor is computed only at the first time step of the event time window and the resulting reference frame is then used for the full event. 
As expected, Fig.~\ref{figure: combined vortex plot}(j) shows that $\mathcal{E}_1$ has the largest share of total enstrophy as the vortices are oriented along the eigenvector $\bm{v}_1$. 
There is a flow of enstrophy from $\mathcal{E}_1$ to $\mathcal{E}_2$ and the time of reconnection is pushed back to $t = 2.4\tau_{\eta}$, significantly later than the previous prediction. 
The transfer of enstrophy offers compelling evidence of reconnection and suggests that the extreme event of downscale $D_l^I$ flux is linked to counter-rotating vortex reconnection. 

\section{Discussion and conclusion}
\label{sec: discussion and conclusion}

Since the pioneering work of \citet{melander1988cut}, the mechanism of reconnection has been well understood. 
Two counter-rotating vortices approach closely, due to self- and mutual induction, cut-and-connect with each other resulting in the formation of bridges and threads. 
For large enough $Re$, these threads, which are themselves counter-rotating, may undergo a further reconnection resulting in the formation of small-scale bridges and threads and the process may continue yet again in a self-similar cascade, each time transferring energy to smaller and smaller scales \citep{hussain2011mechanics, mckeown2018cascade,yao2020physical}. 
It is, therefore, not surprising that vortex reconnection is linked to extreme events of downscale $D_l^I$ flux. 

While this may offer one possible mechanism for the existence of such extreme events below $\eta$, other mechanisms could exist as well. 
For instance, the extreme event H2 (cf. Fig.~\ref{figure: DR plane}(d)) discussed in the Supplementary Section 3 has two vortices spinning around each other suggesting a co-rotating merging event. 
These type of events have been previously observed in \citet{buaria2019extreme} in isotropic turbulence and at large $Re$. They tend to have a three-dimensional character due to the onset of elliptic instability \citep{meunier2005elliptic}. 
This instability, particularly at later stages of merging, is known to generate small-scale structures (cf.~Fig.~22 of \citet{meunier2005elliptic}) which may lead to large values of downscale $D_l^I$ flux.
Further complicated mechanisms may exist when co-rotating and counter-rotating interactions occur simultaneously \citep{leweke2016dynamics}. 
While the existence of scales smaller than $\eta$ is consistent with the previous numerical works \citep{schumacher2010extreme, buaria2019extreme}, our probe of $l = 0.76\eta$ is still close to the Kolmogorov scale. 
Even smaller scales need to be probed before one can conclusively claim the existence of extreme events below $\eta$.

The method discussed in Section \ref{sec: description} is intended to detect only the first-level cut-and-connect of counter-rotating vortex reconnection. 
While it overcomes the subjectivity of defining a vortex, some issues still remain. 
For instance, the event time window and the spatial domain that encompasses both vortices are currently user defined. 
However, one can use a box size and lifetime averaged over the three identified events to find enough similar ones so that one may perform statistics on them. 
Our current efforts are focused in this direction and the results will be part of a future publication. 
The N-tensor is currently applied at the first time step of the event time window. 
However, this was originally conceived to be applicable at all time steps such that even in the presence of a rotating background flow, one may sill be able to detect reconnection. 
In this case, the reference frame needs to be tracked in time and a methodology is being developed. 

\section*{Acknowledgements}

We are grateful to C. Nore and H. Faller for providing us with the DNS simulation of the von K{\'a}rm{\'a}n flow that is used to obtain the visualizations in Fig.~\ref{figure: reconnection}. This research has been funded though the Agence Nationale pour la Recherche, via the grants ANR TILT grant agreement no. ANR-20-CE30-0035, ANR BANG grant agreement no. ANR-22-CE30-0025 and a French government grant managed by the ANR France 2030 under Reference No. ANR-24-RRII-0001. B.M. is supported through the European Union’s Horizon 2020 research and innovation program under the Marie Sklodowska-Curie postdoctoral fellowship grant agreement No. 945298.

\newpage 

\subsection*{Appendix A. The Giant von K{\'a}rm{\'a}n experiment (GvK)}
\label{subsection: GvK}

\begin{figure}[t!]
\centering
\includegraphics[width=0.7\linewidth]{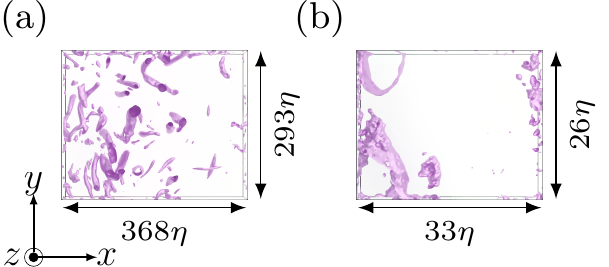}
\caption{Q-criterion structures computed on the measured volume at (a) $Re = 156\, 719$ and (b) $Re = 6\, 269$ showing the different levels of zoom and their respective side lengths in terms of the Kolmogorov length scale.}\label{figure: GvK pictures}
\end{figure}

This is a state of the art experimental facility at SPEC, CEA Paris-Saclay, which improves upon smaller scaled down versions. It consists of a cylindrical tank with $R = 0.5$m and a height of $0.9$m. It is filled with water that is maintained at a constant temperature of $20^{\circ}$C ensuring constant viscosity $\nu$. The flow is forced with with two counter-rotating impellers, each having 8 curved blades. If $Re > 6000$, these impellers drive a quasi-steady overturning and swirling flow which is instantaneously turbulent.  Whether the flow is pushed by the concave (denoted `ANTI') or convex (denoted `CONTRA') side of the blade, there is a strong impact on the turbulent dynamics and statistics \citep{debue2019experimental}. Only the CONTRA cases are considered in this paper.

Although the probed volume is small ($50 \times 40 \times 6$mm$^3$), the actual spatial resolution is determined by the mean inter-particle distance $\Delta x_p = (V/N_p)^{1/3}$ within the volume. 
With constant flow and diagnostic parameters, modifying the rotation frequency (F) directly impacts the Kolmogorov length ($\eta$) and time ($\tau_{\eta}$) scales. For instance, Fig. $\ref{figure: GvK pictures}$ (c, d) shows the size of the same volume in terms of $\eta$ at two different rotational frequencies. Lower rotational frequencies, and hence, lower Reynolds numbers enable the study of scales smaller than $\eta$.

%\subsection*{Particle tracking velocimetry (PTV)}
%\label{subsection: PTV}

The time-resolved velocity fields are obtained with 4D PTV. A high-speed 30mJ pulse Nd:YAG laser is used to highlight a small volume which is positioned at the center of the flow. Four Phantom Micro m340 cameras, each composed of CMOS sensors of $2560 \times 1600$ pixels, were positioned around the setup as illustrated in Fig. 3.13 of \citet{geneste2023evenements}. Spherical polystyrene particles of 5$\mu$m diameter are used for all cases, yielding a Stokes number of $S_{t_{\tau_{\eta}}} \approx 8.1 \times 10^{-5}$. 

The particles are tracked with the `shake-the-box' (STB) algorithm from the commercial Davis10 software \citep{schanz2016shake}. The entire process is described in detail in \citet{cheminet2022eulerian}. Briefly, the first step is calibration with LaVision's 3D calibration plate to take into account image distortion and optical aberrations. This is followed by a self-calibration with particle images where the estimated 3D particle positions are used to reduce calibration errors to below $0.1$ pixels. STB provides particle positions and trajectories of all particles per frame within the volume which contain measurement noise. Two regularized B-spline filters are applied, the first one is a low-pass filter to smooth out the noise along the trajectories and the second one to interpolate the velocity data onto a regular Eulerian field.

This procedure has a filtering effect with a cutoff wavenumber approximated by $k_{\text{PTV}} = 2\pi/\langle d\rangle$, where $\langle d\rangle$ is the mean interparticle distance \citep{geneste2023evenements}. Calculating $D_l^I$ on these velocity fields with a Gaussian filter $\phi^l$ can then have a double filtering effect. This is still meaningful as long as one can ensure that $\phi^l$ does not filter less than the PTV filter. When probing a particular scale of $D_l^I$, the 2 dB cut-off wavenumber of $\phi^l$ is chosen i.e., $l_c = \pi l/\sqrt{30 \ln 2} \approx 0.69 l$. Therefore, the double filtering is meaningful when $l_c \geq 0.5\langle d \rangle$. 
Thus, we note that throughout the paper, when we refer to a scale $l$, we are in fact referring to $l_c$ as it can be directly compared to $\langle d \rangle$.

%% ------------------- BIBLIOGRAPHY ------------------- %
%
\bibliographystyle{unsrtnat} % Bibliography style
\bibliography{pnas-sample.bib} % File where bibliography is stored

\newpage

\subsection*{Supplementary information}
\label{subsection: supplementary}

\subsection*{1. Direct numerical simulation (DNS) of vortex tube reconnection}
\begin{figure}[h!]
    \centering
    \includegraphics[width=\textwidth]{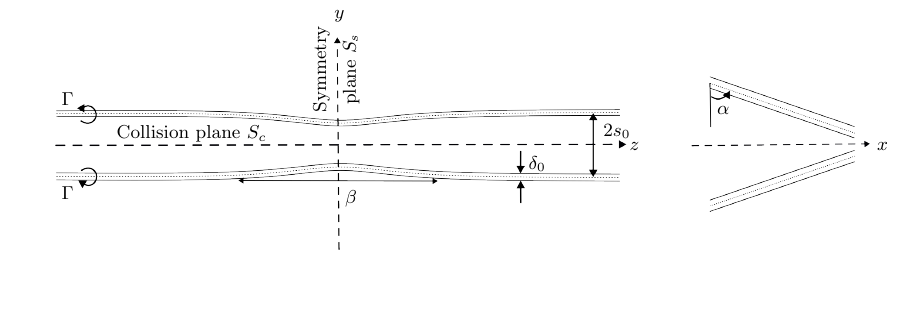}
    \captionsetup{justification=justified}
    \caption{Schematic of the initial configuration used for the DNS simulation.}
    \label{fig: schematic}
\end{figure}

A DNS of two perturbed, antiparallel vortex tubes at a vortex Reynolds number $Re_{\Gamma} = 1000$ is used to support the results presented in the main paper. We choose the hairpin configuration of \citet{hon1991evolution} to generate the vortex tubes with a perturbation,
\begin{equation} \label{eq: initial_curve}
    \mathbf{X}(s) = Ae^{-\beta s^2}\left[(\cos\gamma) \mathbf{e_1} + (\sin\gamma) \mathbf{e_2}\right] + s \, \mathbf{e_3}
\end{equation}
where ($\mathbf{e_1}, \mathbf{e_2}, \mathbf{e_3}$) are unit vectors corresponding to $x, y, z$ directions of a Cartesian coordinate system and $s$ is an arc-length parameter. Perturbation parameters are set as follows: amplitude $A = 0.5$, initial width $\beta = 250$ and inclination angle $\gamma = 60^{\circ}$ with respect to the (y, z) plane. A schematic of the initial configuration is shown in Fig. \ref{fig: schematic}. Since the DNS is carried out with a modified version of the spectral framework PyLogGrid \citep{barral2024pyloggrid}, the initial conditions need to be represented in Fourier space. Following \citet{harikrishnan2025viscous}, the Dirac delta function is used to represent a vortex filament $\mathbf{X}(s)$ in Fourier space. The vorticity is then,
\begin{equation}\label{eq: vorticity_filament}
    \boldsymbol{\omega(x)} = \Gamma\int \delta^{(3)}(\mathbf{x} - \mathbf{X}(s)) \mathbf{T}(s)ds
\end{equation}
where $\mathbf{T}(s) = -2A\beta s \, e^{-\beta s^2}\left[(\cos\gamma) \mathbf{e_1} + (\sin\gamma) \mathbf{e_2}\right] + \mathbf{e_3}$ is the unit tangent vector and $\Gamma$ is the circulation strength. Taking the Fourier transform of the above,
\begin{equation}\label{eq: vorticity_filament_FT}
    \boldsymbol{\hat{\omega}(k)} = \Gamma\int e^{-\mathbf{i} \mathbf{k} \cdot\mathbf{X}(s)} \mathbf{T}(s)ds
\end{equation}
Eq.~\ref{eq: vorticity_filament_FT} is evaluated numerically and multiplied with a 3D Gaussian $e^{-(\|\mathbf{k}\|\delta_0)^2/2}$ to give the filament some thickness. Further multiplication with $e^{ik_y s_0}$ translates the vortex tube in the y-direction with a distance $s_0$. In our case, $\delta_0 = 0.01$, $s_0 = 0.08$ and $\Gamma = \pm 1$ (positive and negative value correspond to different vortex tubes). Multiple vortex tubes are added by superposing the vorticity fields. The initial velocity field is obtained with the Biot-Savart law,
\begin{equation}\label{eq:Biot-Savart}
\mathbf{\hat{u}(k)} = \frac{\mathbf{i}\mathbf{k} \times \boldsymbol{\hat{\omega}(k)}}{\|\mathbf{k}\|^2} \quad \text{for } \mathbf{k}\neq 0
\end{equation}
Simulation is performed with the Fourier-transformed incompressible Navier-Stokes equations,
\begin{align}
\partial_t u_i + \mathbf{i} k_j u_i * u_j &= -\mathbf{i} k_i p - \nu k^2 u_i + f_i, \label{eq: Fourier_INSE_1}\\
\mathbf{i} k_j u_j &= 0, \label{eq: Fourier_INSE_2}
\end{align} 
where $\mathbf{i} = \sqrt{-1}$, $k_i$ is the $i$th component of the d-dimensional wave vector $\bm{k} = (k_1, ..., k_d)$, $p$ is the complex pressure field, $f_i = 0$ denotes the forcing and $\nu$ is the kinematic viscosity. The nonlinear term is convolved in real space and fast Fourier transforms are handled with the parallelized python-SciPy library. The 3/2 padding method is employed for dealiasing. Solutions are advanced in time with an adaptive, exponential time-differencing scheme of \citet{WhalenBrioMoloney2015}. 

The minimum wavenumber is set to $k_{\min} = 2\pi$ such that the computational domain size is $L_x = L_y = L_z = 1$. Following \citet{yao2020physical}, time is non-dimensionalized as $t^* = t\Gamma/2\pi(2s_0)^2$, where $\Gamma$ is the initial circulation strength of one vortex. Finally, the vortex Reynolds number $Re_{\Gamma} = \Gamma/\nu = 1000$ and the grid size is chosen as $129^3$. 

\begin{figure}
    \centering
    \includegraphics[width=0.45\linewidth]{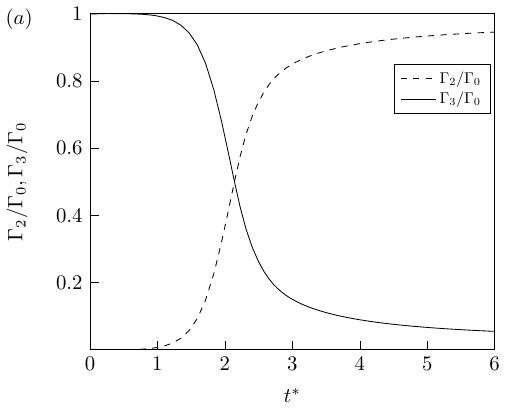}
    \includegraphics[width=0.45\linewidth]{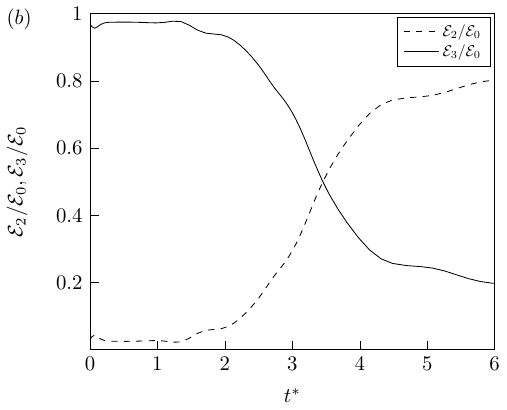}
    \captionsetup{justification=justified}
    \caption{Temporal evolution of (a) circulations $\Gamma_2$ and $\Gamma_3$ and (b) enstrophies $\mathcal{E}_2$ and $\mathcal{E}_3$.}
    \label{fig: transfer}
\end{figure}

The time evolution of reconnection is illustrated in Fig. 1 of the main paper with Q-criterion structures. As pointed out by \citet{yao2020physical}, during reconnection, there is a transfer of circulation from the symmetry to the collision plane. Therefore, temporal evolution of circulation in these two planes can be used as an indicator of reconnection. The circulations are computed along one half-of the $S_s$ and $S_c$ planes as follows,
\begin{align}
\Gamma_2 &= \int_{-L_z/2}^{0} \int_{-L_x/2}^{L_x/2} \omega_2(x, L_y/2, z) \text{d}x \text{d}z \\
\Gamma_3 &= \int_{-L_y/2}^{0} \int_{-L_x/2}^{L_x/2} \omega_3(x, y, L_z/2) \text{d}x \text{d}y
\end{align}
Fig. \ref{fig: transfer}(a) shows the time evolution of $\Gamma_2$ and $\Gamma_3$ normalized by $\Gamma_0 = \Gamma_2 + \Gamma_3$. The circulation remains constant during the inviscid advection phase where the vortex tubes are approaching each other. As the reconnection starts, there is a drop in $\Gamma_3$ in the symmetry plane, and consequently an increase in $\Gamma_2$ in the collision plane. This continues as reconnection progresses and bridge structures are formed in the transverse direction. 

Now, the enstrophy-based indicator is tested. It should be noted that the N-tensor is unnecessary as the vortex tubes are oriented along the $z$-axis and the reference frame is known. Since $\mathcal{E}_3$ and $\mathcal{E}_2$ are the relevent components to study the enstrophy transfer, their time evolution is calculated, normalized by $\mathcal{E}_0 = \mathcal{E}_2 + \mathcal{E}_3$ and plotted in Fig. \ref{fig: transfer}(b). Similar to the circulation plot, a transfer of componentwise enstrophy from $\mathcal{E}_3$ to $\mathcal{E}_2$ is evident. It is interesting to observe that this is less smooth than the circulation plot. Since componentwise enstrophy is calculated over the full domain, it can be influenced by the presence of additional structures (e.g., threads). 

\subsection*{2. Diagnostics for event L1}

\begin{figure}
\centering
\includegraphics[width=\textwidth]{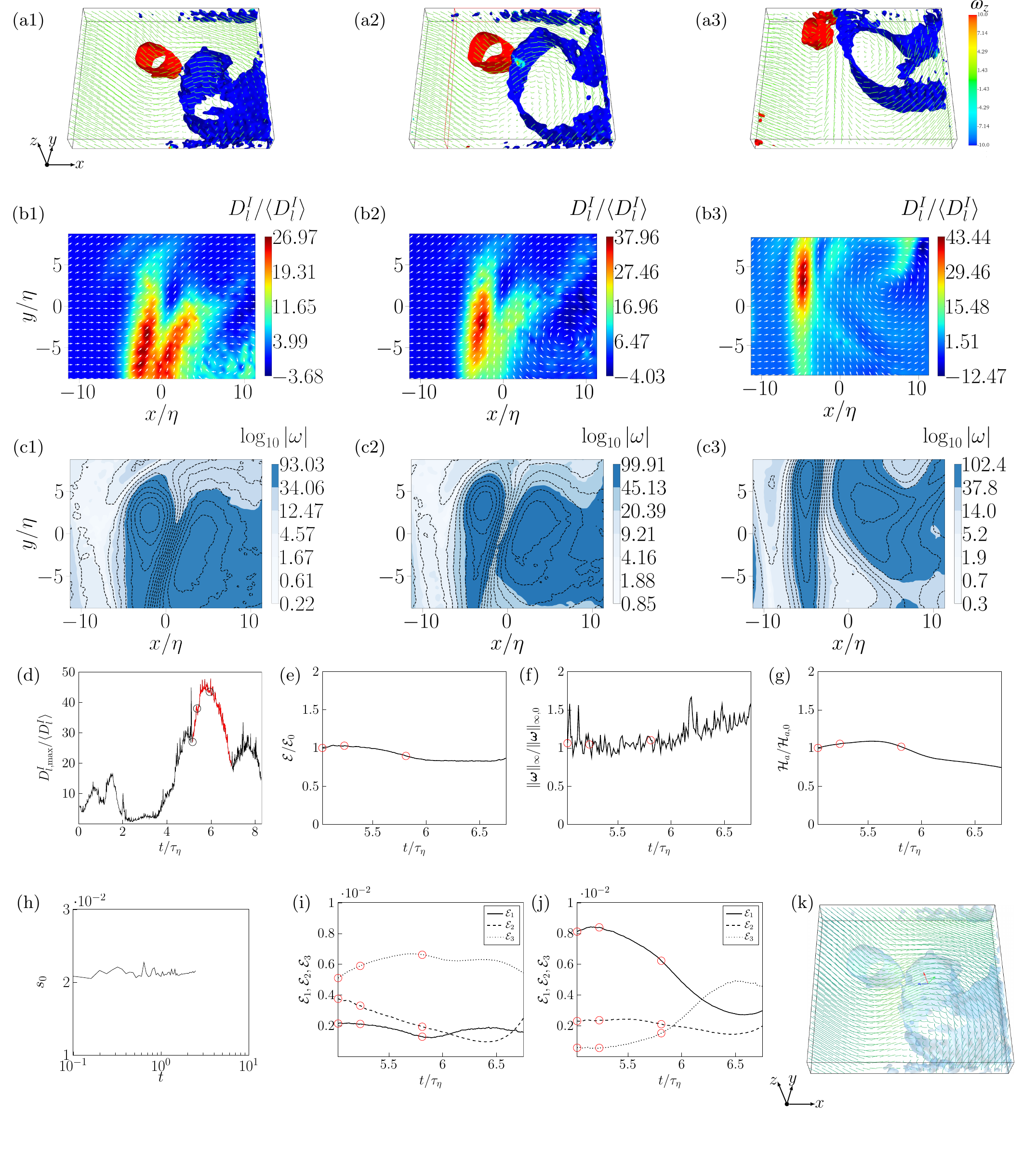}
\captionsetup{justification=justified}
\caption{(a1-a3) Isosurfaces of the Q-criterion thresholded at Q $> 0.12$ Q$_{\max}$ and shaded with axial vorticity ($\omega_z$). Velocity streamlines at mid-z height are shown. Horizontal slices of (b1-b3) mean-normalized $D_l^I$ and (c1-c3) vorticity magnitude are plotted at slices where the mean-normalized $D_l^I$ reaches an instantaneous maximum. All the above are plotted at times $t /\tau_{\eta} = 5.16, 5.37, 5.94$. These times are indicated in (e-g, i-j) with unfilled red markers. (d) Full time series of mean-normalized $D_l^I$ for the run. Temporal evolution of (e) enstrophy, (f) maximum norm of vorticity and (g) net absolute helicity, all of which are normalized by their initial value. Behavior of (h) separation distance, component-wise enstrophy computed (i) without and (j) with the new reference frame obtained from the N-tensor. This reference frame is indicated in (k) and the red arrow identifies the direction associated with the eigenvector $v_{\lambda_{\max}}$.}
\label{fig: event_L1}
\end{figure}

Similar to event L2 described in the main paper, it can be seen from Figs. \ref{fig: event_L1}(a1-a3) that two large Q-criterion structures exist. As they are roughly oriented in the $z$-direction, coloring them by axial vorticity ($\omega_z$) reveals that they are counter-rotating. It is interesting to note that the Q-criterion structure colored red, appears elongated in the vorticity magnitude plots (cf. Figs. \ref{fig: event_L1}(c1-c3)). Since it is well-known that vorticity magnitude cannot distinguish between spin and shear as they both possess nonzero vorticity \citep{lugt1979dilemma}, the difference between the visualizations suggests that the elongated regions are shearing. This could be indicative of the core-flattening effect that eventually leads to the formation of vortex sheets at the point of contact during reconnection (cf. figure 6 of \citet{yao2022vortex}). Figs. \ref{fig: event_L1}(b1-b3) suggest that large values of mean-normalized $D_l^I$ are concentrated around these shearing regions. 

Within a small box of size $20.7\eta \times 19.9\eta \times 3.2\eta$ (colored red in Fig. \ref{fig: event_L1}(a2)) and the event time window (also colored red in Fig. \ref{fig: event_L1}(d)), it can be seen that there is a minor rise and fall of total enstrophy, supremum norm of vorticity and absolute helicity (Figs. \ref{fig: event_L1}(e-g)). As seen from Fig. \ref{fig: event_L1}(h), the separation distance scaling could not be obtained as the vortices are already too close to each other at the beginning of the event time window. The analysis could not be improved even if the tracking was started at an earlier time. The enstrophy-based criterion detects a reconnection only when the reference frame suggested by N-tensor is used (cf. Figs. \ref{fig: event_L1}(i, j)), thereby highlighting the importance of identifying the correct reference frame. 

\subsection*{3. Diagnostics for events H1 and H2}

\begin{figure}
\centering
\includegraphics[width=0.9\textwidth]{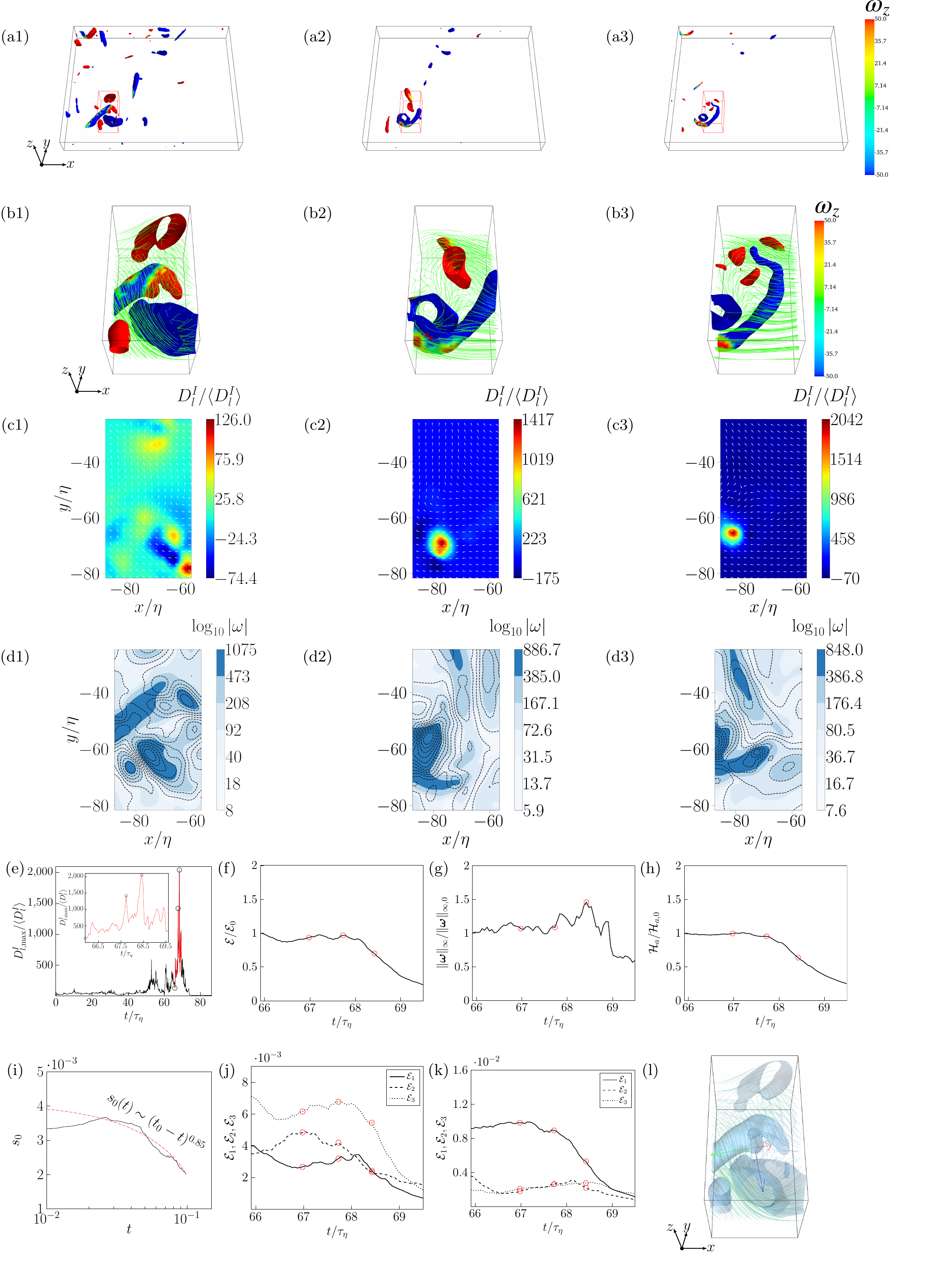}
\captionsetup{justification=justified}
\caption{(a1-a3) Isosurfaces of the Q-criterion thresholded at Q $> 0.05$ Q$_{\max}$ and shaded with axial vorticity ($\omega_z$). (b1-b3) The isosurfaces are shown within the red box in (a1-a3). Velocity streamlines at mid-z height are shown. Horizontal slices of (c1-c3) mean-normalized $D_l^I$ and (d1-d3) vorticity magnitude are plotted at slices where the mean-normalized $D_l^I$ reaches an instantaneous maximum. All the above are plotted at times $t /\tau_{\eta} = 65.91, 67.73, 68.43$. These times are indicated in (f-h, j-k) with unfilled red markers. (e) Full time series of mean-normalized $D_l^I$ for the run. Temporal evolution of (f) enstrophy, (g) maximum norm of vorticity and (h) net absolute helicity, all of which are normalized by their initial value. Behavior of (i) separation distance, component-wise enstrophy computed (j) without and (k) with the new reference frame obtained from the N-tensor. This reference frame is indicated in (k) and the red arrow identifies the direction associated with the eigenvector $v_{\lambda_{\max}}$.}
\label{fig: event_H1}
\end{figure}

\begin{figure}
\centering
\includegraphics[width=0.9\textwidth]{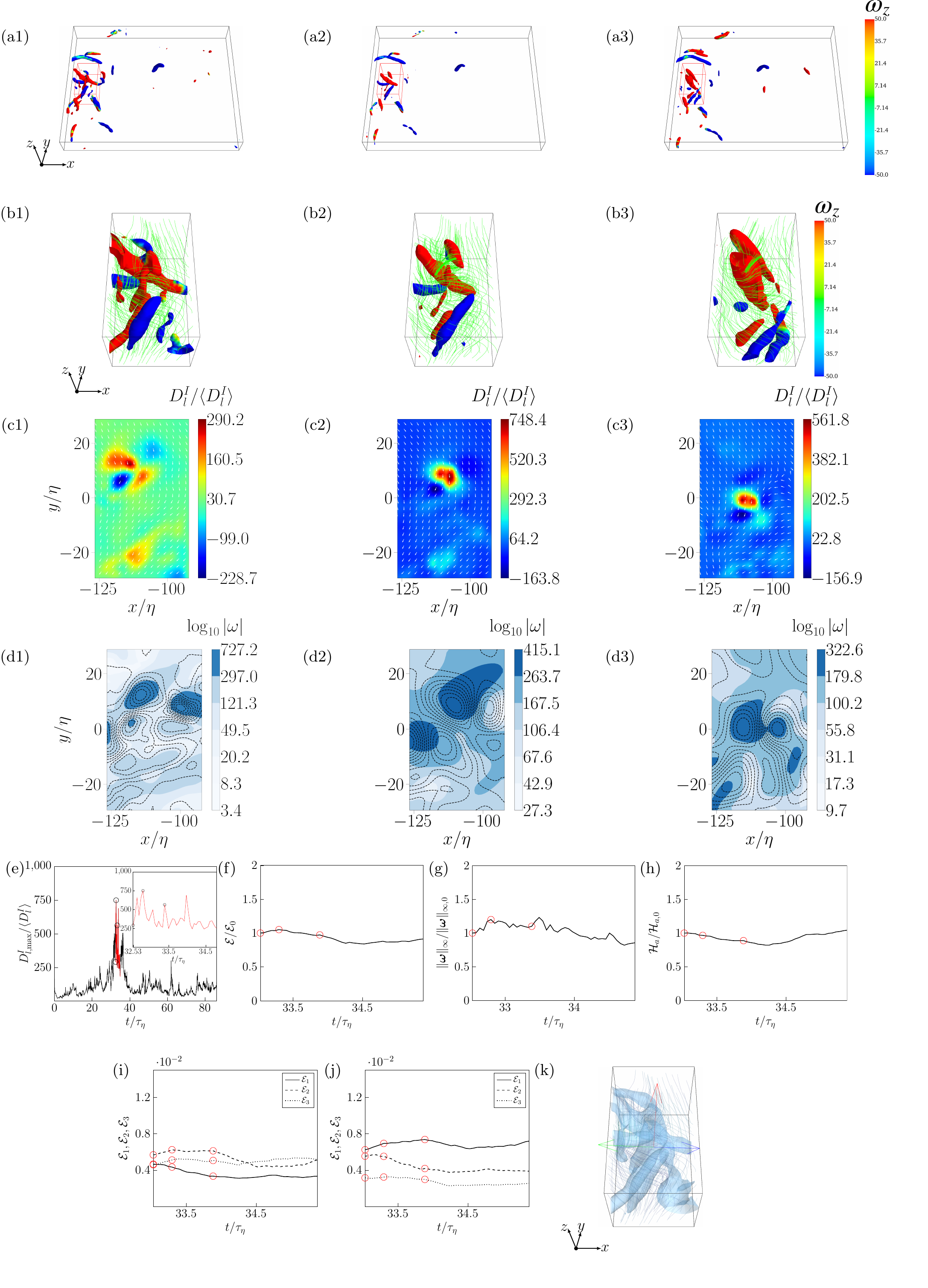}
\captionsetup{justification=justified}
\caption{(a1-a3) Isosurfaces of the Q-criterion thresholded at Q $> 0.05$ Q$_{\max}$ and shaded with axial vorticity ($\omega_z$). (b1-b3) The isosurfaces are shown within the red box in (a1-a3). Velocity streamlines at mid-z height are shown. Horizontal slices of (c1-c3) mean-normalized $D_l^I$ and (d1-d3) vorticity magnitude are plotted at slices where the mean-normalized $D_l^I$ reaches an instantaneous maximum. All the above are plotted at times $t /\tau_{\eta} = 32.53, 32.79, 33.38$. These times are indicated in (f-h, j-k) with unfilled red markers. (e) Full time series of mean-normalized $D_l^I$ for the run. Temporal evolution of (f) enstrophy, (g) maximum norm of vorticity and (h) net absolute helicity, all of which are normalized by their initial value. Behavior of component-wise enstrophy computed (i) without and (j) with the new reference frame obtained from the N-tensor. This reference frame is indicated in (k) and the red arrow identifies the direction associated with the eigenvector $v_{\lambda_{\max}}$.}
\label{fig: event_H2}
\end{figure}

This section discusses results from the high Reynolds number case C2. It should be noted that the scale probed for $D_l^I$ is at $l = 8.52\eta$. A major difference compared to case C1 is that plenty of vortex structures exist within the domain. Consequently, the $z$-direction extent is an order of magnitude larger at $35.5\eta$. 

The event H1 (Fig. \ref{fig: event_H1}) qualifies as a typical `extreme event' where the spatial maxima of $D_l^I$ exceeds thousands of its mean value in strength. The Q-criterion visualizations from Figs. \ref{fig: event_H1}(a1-a3, b1-b3) reveal a counter-rotating vortex pair within a small box of size $35.5\eta \times 59\eta \times 35.5\eta$. Similar to events L1 and L2, the vortices are roughly oriented in the $z$-direction. Applying the diagnostics within the box and the event time window (highlighted red in Fig. \ref{fig: event_H1}(e)) reveals that the total enstrophy, supremum norm of vorticity and the absolute helicity indicate a minor rise and fall at a later time (around 67-68$\tau_{\eta}$). This is consistent with the enstrophy-based detection criterion where a flow of enstrophy is observed nearly at the end of the event time window (cf. Figs. \ref{fig: event_H1}(j, k))). By tracking the vortices in time, we obtain a separation distance scaling $s_0 \sim (t_0 - t)^{0.85}$ which is similar to the event L2. 

Unlike the previously discussed events where the enstrophy-based detection is successful, the event H2 is included as an example where an alternate path to generate large values of $D_l^I$ flux can exist. Since the vortices are roughly oriented in the $z$-direction, coloring with axial vorticity reveals that the two vortices are co-rotating (cf. Fig. \ref{fig: event_H2}(a3, b3)). While the enstrophy-based detection criterion without the N-tensor (cf. Fig \ref{fig: event_H2}(i)) reveals a flow of enstrophy from $\mathcal{E}_2$ to $\mathcal{E}_3$, this is a false positive as the orientation of the vortices suggests that $\mathcal{E}_3$ should have been the dominant component of enstrophy and an opposite flow from $\mathcal{E}_3$ to $\mathcal{E}_2$ should have occurred (see Fig. \ref{fig: event_H1}(j) and Fig. 5(i) in the main paper where the proper flow of enstrophy was observed). In the correct reference frame, no transfer of enstrophy among its components is detected (cf. Fig \ref{fig: event_H2}(j)). This is expected since the enstrophy-based detection criterion is designed to detect a first-level cut and connect which is absent in a co-rotating merging event.

%%% Each figure should be on its own page

\end{document}